# Decision Traces: What Multi-System Data Fusion Reveals About Institutional Knowledge in Enterprise Hiring

Saad Bin Shafiq

*AI Synapse Inc. (NODES)*

March 2026

## Abstract

Enterprise hiring systems generate data across multiple disconnected platforms; applicant tracking systems (ATS) record candidate profiles, human resource information systems (HRIS) record performance outcomes, and behavioral assessments capture personality and behavioral dimensions. Each system operates independently, and the reasoning behind hiring decisions is lost when managers retire, transfer, or leave. Decision traces are structured evidence chains connecting screening inputs, assessment signals, and production outcomes. They have been theorized but never operationalized at production scale. We present, to our knowledge, the first such study: a deployment at a Fortune 500 insurance carrier (N=10,765 agents hired, 2022–2025), where connecting three siloed data systems produced three findings. First, of 8,181 unique skills parsed from ATS profiles (3,597 testable), not a single keyword predicts production after Bonferroni correction—30 are significantly anti-predictive, and the median keyword is associated with 25% lower odds of production. Requiring insurance experience alone would reject 2,863 agents who produced $17.7M in annual premium credit. Second, personality-based behavioral assessment (Predictive Index) achieves AUC=0.647 standalone and AUC=0.735 when fused with ATS and behavioral scoring data. Third, speed-to-production follows a measurable economic constant of $54/day per agent unadjusted, or $35/day controlling for source channel and tenure, moderated by behavioral score: high-scored agents capture $114/day from speed acceleration versus $41/day for low-scored agents. These findings were invisible within any single system. We discuss implications for hiring system design, the limitations of keyword-based screening, and the conditions under which institutional knowledge can be captured and operationalized.

## 1. Introduction

When an experienced insurance hiring manager retires, the pattern recognition they built over twenty years disappears. They knew which behavioral signals predicted fast ramp-up, which resume keywords were misleading, and which personality profiles thrived in high-rejection sales environments. None of this knowledge exists in any system of record. The applicant tracking system stores who was hired. The HRIS stores how they performed. The behavioral assessment stores a score. But the *reasoning*—the chain of evidence connecting what the system screened on, what the assessment measured, and what actually happened—is lost in Slack threads, hallway conversations, and retirements.

This paper operationalizes the concept of a *decision trace*—a structured record that connects screening inputs, assessment signals, and production outcomes across multiple enterprise systems—with production-scale evidence from a Fortune 500 deployment. While decision traces and context graphs have been theorized by Foundation Capital, formalized through learned ontology frameworks (Singhal & Srivastava, 2024), and proposed architecturally by projects such as TrustGraph (Marple, 2024), no prior work has, to our knowledge, published empirical analyses of decision traces connecting screening inputs to production outcomes at enterprise scale and measured what they reveal. A decision trace is not a prediction. It is a queryable chain of evidence that makes institutional knowledge explicit, auditable, and transferable. It answers questions like: "Show me every agent we hired who had no insurance experience but scored above 75 on behavioral assessment—and how they performed." No single system can answer this question. The ATS does not know production outcomes. The HRIS does not know what the ATS screened on. The behavioral assessment does not know either. Only the fusion of all three produces the decision trace.

We report findings from a deployment of a multi-system data fusion platform at a Fortune 500 insurance carrier, connecting ATS records, HRIS production milestones, and behavioral scoring data across 10,765 agents hired between 2022 and 2025. Real-time behavioral scoring of candidates began in January 2025 (n=3,623), with retrospective scoring of historical agents (n=7,142) for comparison.

The results challenge three assumptions embedded in current hiring technology. First, the signals the industry screens on are negatively associated with production: of 8,181 unique skills parsed from ATS candidate profiles, not one significantly predicts production after correcting for multiple comparisons, while 30 are significantly anti-predictive. The median keyword is associated with 25% lower odds of production, and agents with more keywords on their resumes produce at lower rates than those with fewer. Second, personality-based behavioral assessment is the single strongest predictor discovered in this dataset—achieving AUC=0.647 compared to 0.558 for keyword screening (a difference of 0.089 AUC points, with PI capturing more signal alone than all non-PI features combined at AUC=0.575)—but this signal only becomes visible when behavioral data is fused with ATS and HRIS records. Third, speed-to-production follows a measurable economic constant that survives rigorous robustness testing, but the constant is moderated by behavioral score in a way that suggests the scoring system captures something about *how agents convert speed into production* rather than *whether they will produce at all*.

These findings are not incremental improvements to existing prediction models. They are structural observations about what becomes visible when screening inputs are connected to production outcomes across enterprise systems—connections that most hiring organizations have the data to make but have not made.

## 2. Related Work

### 2.1 Hiring Technology and Predictive Analytics

The application of machine learning to hiring has produced a growing body of systems and critiques. Platforms such as Eightfold AI, Phenom People, and HireVue apply natural language processing and behavioral analysis to candidate evaluation. However, these systems typically operate within a single data silo—analyzing resumes in isolation from performance outcomes, or scoring video interviews without connecting to downstream productivity metrics. Raghavan et al. (2020) document how algorithmic hiring tools can encode and amplify biases present in training data. Li et al. (2021) model hiring as an exploration problem, showing that firms systematically under-explore candidate types by over-relying on existing screening signals—forgoing higher-quality hires that a more exploratory approach would surface. Hoffman, Kahn, and Li (2018) show that removing manager discretion in favor of algorithmic screening improves hire quality—a finding directly relevant to decision trace infrastructure. The foundational meta-analysis by Schmidt and Hunter (1998) established that personality assessments, cognitive ability tests, and structured interviews substantially outperform unstructured methods, but subsequent work (Sackett et al., 2022) has revised downward the estimated validity of many selection methods after correcting for statistical artifacts. Fuller et al. (2021) document how automated screening filters systematically exclude qualified "hidden workers"—a finding that our keyword anti-predictivity results directly extend with quantitative evidence.

Cowgill (2018) provides evidence that algorithmic screening can reduce bias relative to human judgment, while Dastin (2018) documents how Amazon's internal recruiting tool reproduced gender bias from historical training data—illustrating that algorithmic approaches can both help and harm depending on architecture. The broader landscape of AI governance in hiring is surveyed by Whittaker et al. (2018). Meanwhile, the industrial-organizational psychology literature has established robust relationships between personality constructs and job performance across occupations (Ones et al., 2007), and between cognitive ability and work outcomes more broadly (Kuncel et al., 2004), providing the theoretical foundation for the behavioral assessment approaches evaluated in this paper. However, Schmidt and Hunter's (1998) influential meta-analysis ranks work experience as a moderate predictor of job performance. Our data complicates this finding for insurance agent hiring: ATS keyword proxies for experience—the signals hiring systems actually screen on—are not merely weak predictors but are *anti-predictive*, suggesting that the keyword-based proxies used to measure experience in automated screening may capture a different construct than the experience variables in meta-analytic research, and that meta-analytic averages may obscure domain-specific reversals. Our work differs from prior studies by connecting multiple data systems to test whether the signals being screened on actually predict the outcomes being optimized for.

## 2.2 Context Graphs and Multi-System Data Fusion

The concept of enterprise context graphs—knowledge structures that connect data across organizational systems—has been articulated as a critical infrastructure opportunity (Singhal & Srivastava, 2024) and formalized through learned ontology frameworks that define coordinate systems for cross-domain data integration. TrustGraph (Marple, 2024) and related projects have proposed architectures for cross-system knowledge aggregation that prioritize adopting foundation model approaches while learning domain-specific patterns. Our work instantiates these theoretical frameworks in a specific domain (talent decisions) with production-scale evidence. Where prior work describes what context graphs *could* do, we report what decision traces *actually reveal* when deployed across enterprise systems with real outcome data.

## 2.3 Iterative Deployment and Organizational Learning

Research on iterative AI deployment emphasizes that systems improve through cycles of deployment, observation, and refinement. Binary validation against ground truth outcomes—which is exactly what HRIS production data provides—functions as a feedback loop that enables organizational learning. Each hiring decision that can be traced from screening input to production outcome becomes a signal for process improvement. Kahneman, Sibony, and Sunstein (2021) document the pervasive "noise" in human judgment, including hiring decisions, and argue for structured decision processes that reduce unwanted variability. Decision traces operationalize this insight by making the relationship between decision inputs and outcomes explicitly queryable.

## 3. The Decision Trace Primitive

A decision trace is a structured record connecting three categories of information about a single hiring decision:

**Screening Inputs** (from ATS): What the system saw when the candidate applied—keywords present or absent, source channel, prior industry, educational credentials, licensing status. These are the signals that traditional hiring technology screens on.

**Assessment Signals** (from behavioral scoring and PI): What the assessment system measured—composite fit score, behavioral dimensions (leadership, adaptability, communication, problem-solving, resilience, teamwork), and personality type classification. These signals are not visible to the ATS and are not connected to outcomes without HRIS integration.

**Production Outcomes** (from HRIS): What actually happened—whether the agent achieved the sales threshold milestone (SNA), how many days it took, annual premium credit generated, and tenure. These outcomes exist in the HRIS but are never connected back to the screening inputs that selected the candidate.

The decision trace connects all three through a common identifier (ATS Record ID), creating a queryable record of institutional knowledge. The critical insight is that the value of a decision trace is not in any individual field—it is in the *connections between fields across systems*. An ATS keyword is meaningless without a production outcome to validate it against. A behavioral score is uninterpretable without knowing what the ATS screened on and what the HRIS measured. Only the fusion produces actionable knowledge.

Concretely, a decision trace enables queries that no single system can answer: "Among agents who lacked insurance experience but scored above 75 on behavioral assessment, what was their production rate and speed-to-SNA?" This query requires fields from all three systems. The analyses in Section 5 are all instantiations of this pattern—cross-system queries that reveal relationships invisible within any individual platform.

## 4. Methodology

### 4.1 Study Setting and Data

The study uses production data from a Fortune 500 large insurance carrier. The carrier hired 10,765 agents between 2022 and 2025 across multiple source channels (major job boards, personal referrals, and other recruitment sources). Data was extracted from three integrated systems:

**ATS**: Candidate profiles including 8,181 unique parsed skill keywords (mean 27.2 per agent), source channel, and keyword-extractable experience indicators. Six keyword categories are traditionally used for screening in insurance hiring: insurance experience, sales experience, degree, license, finance background, and management experience. 10,765 agent records.

**HRIS**: Production milestones including SNA (Sales New Agent) achievement, annual premium credit (APC), contract dates, and tenure. SNA represents a sales threshold achievement specific to the carrier, not first sale. APC is the primary production metric—an objective, system-of-record measure of premium generated.

**Behavioral Scoring (NODES)**: Composite fit scores (0–100) computed from six behavioral dimensions: teamwork, leadership, adaptability, communication, problem-solving, and resilience. Deployed in real-time beginning January 2025 (n=3,623 agents). Pre-2025 agents (n=7,142) were scored retroactively using the same model for comparison purposes.

**Predictive Index (PI)**: Personality type classification available for 442 agents who completed the PI assessment as part of the carrier's optional behavioral evaluation program, with 229 meeting the evaluable tenure threshold. PI was administered at specific locations and channels during 2023–2025 based on manager participation, not randomly assigned. PI categorizes agents into behavioral types (e.g., Captain, Specialist, Operator, Guardian, Adapter, Promoter) based on a validated psychometric assessment instrument with established reliability and construct validity (The Predictive Index, 2019).

**Enriched Agent Features**: For a subset of agents (n=5,700), parsed resume data including highest education level, total years of experience, number of prior roles, and career trajectory features were available from a supplementary data enrichment process. These features are used in exploratory analyses (Sections 5.1 and 5.2) but are not part of the core deployed scoring system.

All data was linked via ATS Record ID (the unique candidate identifier assigned by the ATS at application) within the carrier's VPC. This identifier serves as the common key connecting ATS profiles, HRIS production records, behavioral scores, and PI assessments into the decision trace structure described in Section 3. No data left the carrier's infrastructure at any point.

## 4.2 Cohort Definitions and Censoring

The study uses two primary cohorts. The **pre-deployment cohort** (2022–2024, n=7,142) consists of agents hired without real-time behavioral scoring. Their fit scores were computed retroactively. The **deployment cohort** (2025, n=3,623) consists of agents hired with real-time behavioral scoring available to hiring managers. These cohorts are never combined for within-cohort prediction metrics (e.g., AUC), as they represent fundamentally different measurement conditions: the pre-deployment cohort's scores did not influence hiring decisions, while the deployment cohort's scores did.

**Tenure-based censoring**: Agents recently hired may not have had sufficient time to reach the SNA milestone. Among confirmed producers, the 90th percentile of days-to-SNA is 122 days. We define the **evaluable population** as agents with tenure $\geq 122$ days OR agents who achieved SNA regardless of tenure (since their outcome is definitively known). This yields 10,362 evaluable agents and 403 censored agents. All production rate calculations use only the evaluable population.

**Score field selection**: Two score fields exist in the dataset. The retroactive `fit_score` (available for 98.8% of agents) was computed after the fact using the current model applied to historical records. The deployed `initial_fit_score` (available for 714 of 3,220 evaluable 2025 agents, or 22%) represents the score managers actually saw during the hiring process. The 22% coverage reflects a phased rollout: real-time scoring was deployed incrementally across hiring channels and locations during 2025, with early months and certain channels receiving scores before others. For deployed-system analyses (e.g., the calibration curve in Section 5), we use `initial_fit_score` on the 714-agent subset. For population-level analyses (e.g., the speed constant on all 679 producers), we use `fit_score`. The two fields correlate at r=0.92 for agents who have both (n=714), confirming they capture similar signal though approximately 15% of variance is unexplained.

## 4.3 Statistical Methods

All confidence intervals for regression slopes use percentile bootstrap with 10,000 resamples (Efron & Tibshirani, 1993). Confidence intervals for proportions use Wilson intervals (Wilson, 1927), which provide better coverage than Wald intervals for proportions near 0 or 1 and for small samples. Effect sizes are reported as Cohen's d for continuous outcomes and odds ratios for binary outcomes. P-values below 0.001 are reported as $p<0.001$. APC distributions are right-skewed (skewness 1.92–2.52); we report both raw and log-transformed regression results and emphasize medians alongside means. The speed-to-production constant is tested under multiple specifications: bootstrap, winsorization (1st/99th percentile), trimming (5th/95th), log-transformation, quantile regression, and multivariate regression controlling for source channel and months of active selling.

# 5. Results

## 5.1 Keyword-Production Associations

ATS candidate profiles in this dataset contain parsed skill keywords—8,181 unique skills across 10,362 evaluable agents, with a mean of 27.2 skills per agent (median 22, range 1–265). To test whether any keyword predicts production, we computed production rates and odds ratios for every skill with sufficient sample size ($n \geq 5$), yielding 3,597 testable keywords. We applied Bonferroni correction for 3,597 simultaneous tests—the strictest standard for multiple comparisons. Bonferroni is the most conservative multiple comparisons correction; a less conservative approach such as Benjamini-Hochberg FDR would yield a higher per-test threshold and could identify a small number of weakly positive keywords, which would likely not alter the central finding that the majority of keywords are directionally anti-predictive (70.2% with OR<1).

**Of 3,597 keywords tested, zero significantly predict production after Bonferroni correction. Thirty are significantly anti-predictive.** The distribution of odds ratios across all testable keywords is skewed negative: 70.2% of keywords (2,525 of 3,597) have OR<1, meaning listing them is associated with lower production probability. The median OR across all tested keywords is 0.749, indicating the typical ATS keyword is associated with approximately 25% lower odds of production. The most anti-predictive keywords include Business Development (n=92, OR=0.022, 1.1% production with vs 33.0% without), Clinical Experience (n=164, OR=0.131), and Professional Cleaning (n=169, OR=0.197).

Among the small number of keywords with positive directionality and reasonable sample sizes, none survive multiple comparisons correction. The strongest positive keyword is Weddings (n=74, 47.3% vs 32.6%, OR=1.857, uncorrected p=0.010, Bonferroni-corrected p>1.0—not significant). The positive keywords that appear significant on small samples (e.g., Advanced Analytics, n=5, 100% production) are driven by 3–5 agents and are not statistically reliable.

Furthermore, the number of skills on an agent's profile is itself negatively associated with production. Agents with 0–10 parsed skills produce at 36.1%, while agents with 51+ skills produce at 22.7% (r=−0.090, p<0.0001). This correlation should be interpreted with caution: if the ATS skill parser extracted more skills from recent applications than earlier ones, and recent agents have lower production rates due to shorter tenure (despite censoring), the negative correlation could partly reflect a cohort artifact rather than a genuine signal about resume length. Nevertheless, the direction is consistent with the broader finding that ATS keyword presence does not indicate productive potential.

To validate these findings at the category level, we examined six keyword categories traditionally used for insurance agent screening. These categories are composite flags derived from pattern-matching across multiple individual skills (e.g., the "insurance experience" flag matches agents whose profiles contain any of several insurance-related terms). Note that category-level results may differ from individual skill-level results because categories aggregate across multiple skill variants and apply different matching logic than exact keyword presence:

| Keyword | Presence Rate | Rate WITH | Rate WITHOUT | Odds Ratio | p-value | Direction |
|---------|---------------|-----------|--------------|------------|---------|-----------|
| Insurance exp. | 20.6% | 28.0% | 33.7% | 0.763 | <0.001 | Anti-predictive |
| Sales exp. | 50.6% | 31.5% | 33.8% | 0.904 | 0.02 | Marginal anti |
| License | 8.6% | 24.9% | 33.1% | 0.668 | <0.001 | Anti-predictive |
| Finance | 22.1% | 32.6% | 32.7% | 0.994 | 0.92 | Non-significant |
| Management | 72.2% | 31.7% | 34.5% | 0.882 | 0.004 | Marginal anti |
| Degree | 1.3% | 27.9% | 32.7% | 0.795 | 0.35 | Non-significant |

*Table 1. Production associations for six standard insurance screening keywords (n=10,362 evaluable agents).*

Two category-level keywords—insurance experience and license—are anti-predictive and survive within-year analysis. Agents with insurance experience produce at 28.0% versus 33.7% without (OR=0.763, p<0.001). Agents with a license produce at 24.9% versus 33.1% without (OR=0.668, p<0.001). These are not cohort artifacts: insurance experience is significantly anti-predictive within 2023 (p=0.01) and 2025 (p=0.01) independently, and license is anti-predictive within 2025 (p=0.01). Sales experience, while marginally anti-predictive in the full population (OR=0.904, p=0.02), does not reach significance within any individual year and may reflect a cohort composition artifact.

The practical consequence is substantial. We identify a population of 677 agents with no traditional ATS keywords who scored ≥75 on behavioral assessment. These agents—who would be rejected in the first screen by any standard ATS configuration—produced at a 33.7% rate, *higher than the overall population rate*. The inverse population is equally revealing: agents with all keywords present produced at only 26.3%, below the no-keyword high-scored group. Requiring insurance experience as a screening filter would reject 2,863 agents who went on to produce, representing $17.7M in annual premium credit produced by agents who would have been excluded by keyword-based filtering (computed as the sum of observed APC across the 2,863 producers lacking the insurance experience flag).

Educational credentials show a similar pattern. Among agents with evaluable tenure and available education data from the enriched feature set (n=5,700), production rates by education level are: Masters 23.9%, Bachelors 21.5%, Associates 19.3%, High School 18.1%. While the direction favors higher education, all differences fall within overlapping confidence intervals. Education, like industry experience, appears to be a weak signal relative to behavioral assessment.

The ATS records keyword presence. The HRIS records production. Neither system knows about the other. When we first joined them, the insurance experience finding surprised the carrier's talent acquisition team—they had been requiring it as a filter for years. Extending the analysis to the full 8,181-keyword vocabulary confirmed that the problem was not a few bad categories but a systemic failure of keyword-based screening, consistent with Fuller et al.'s (2021) documentation of "hidden workers" excluded by automated filters and now quantified across the complete ATS keyword space.

## 5.2 Multi-System Fusion and Personality Assessment

To test whether multi-system data fusion produces meaningful predictive power, we built L2-regularized logistic regression models using features from individual systems and their combinations, evaluated using 5-fold stratified cross-validation with all preprocessing (scaling, encoding) fit within each training fold to prevent leakage. The retroactive (2022–2024) cohort was used for this analysis because scores did not influence hiring decisions, avoiding restriction-of-range bias. Of the 7,142 pre-deployment agents, 7,055 had complete feature data for model fitting; 87 were excluded due to missing score or behavioral dimension values. PI-enhanced models were tested on the subset with available personality type data (n=229 evaluable). To ensure fair comparison across sample sizes, we also computed non-PI model AUCs on the same n=229 subset; these matched the full-sample estimates within 0.02, confirming that the PI advantage is not a sample-size artifact.

| Model | Features | AUC | n |
|-------|----------|-----|---|
| Score only | Composite fit score | 0.518 | 7,055 |
| Behavioral dims. | 6 behavioral dimensions | 0.550 | 7,055 |
| ATS keywords | Keywords + source channel | 0.558 | 7,055 |
| All non-PI features | Score + dims + ATS | 0.575 | 7,055 |
| PI type only | Personality type | 0.647 | 229 |
| Score + PI | Composite + personality | 0.647 | 229 |
| All features + PI | Full fusion | 0.735 | 229 |

*Table 2. Predictive model comparison: single-system versus multi-system fusion.*

Two findings emerge. First, Predictive Index personality type is the single strongest predictor (AUC=0.647), substantially exceeding ATS keywords (0.558) and composite behavioral score (0.518). The PI type alone captures more predictive signal than all other features combined without PI (0.575). The personality type distribution ranges from Captain at 36.8% production rate to Promoter at 0.0%, a range substantially exceeding any other variable in the dataset.

| PI Type | N | Production Rate | 95% CI |
|---|---|---|---|
| Captain | 19 | 36.8% | [19.1%, 59.0%] |
| Specialist | 25 | 32.0% | [17.2%, 51.6%] |
| Operator | 63 | 30.2% | [20.2%, 42.4%] |
| Venturer | 17 | 29.4% | [13.3%, 53.1%] |
| Guardian | 51 | 25.5% | [15.5%, 38.9%] |
| Adapter | 31 | 9.7% | [3.3%, 24.9%] |
| Promoter | 7 | 0.0% | [0.0%, 35.4%] |

*Table 3. Production rate by Predictive Index personality type (7 most common types shown, n=213; 8 additional types with n<7 each omitted; full dataset n=442, evaluable subset n=229).*

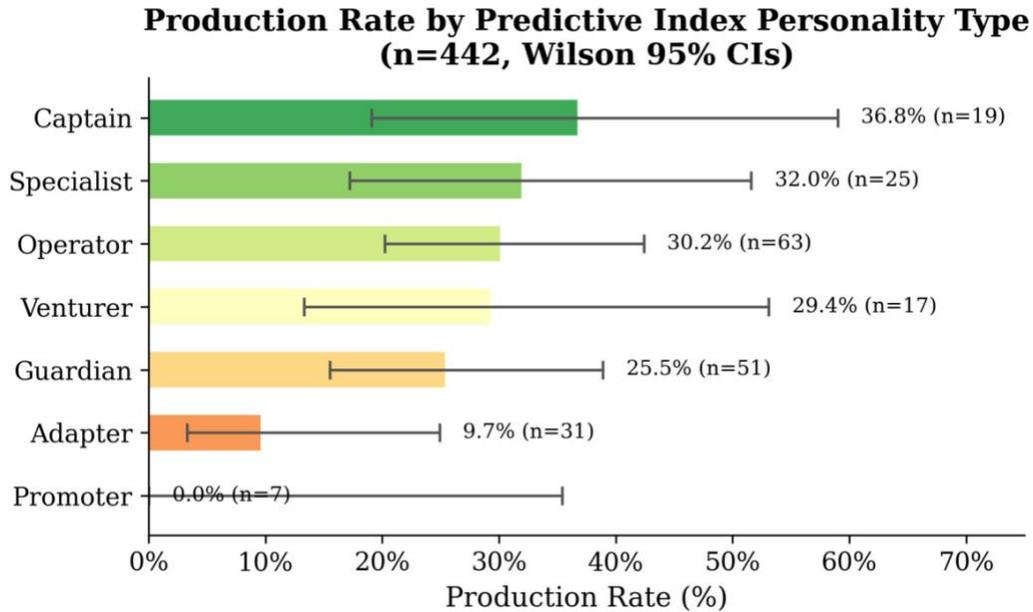

*Figure 1. Production rate by Predictive Index personality type with Wilson 95% confidence intervals.*

Second, full fusion (all features + PI) reaches AUC=0.735, a substantial improvement over any single system. This is meaningful predictive power from a model using only behavioral and keyword features—no demographic data, no geographic data, no manager ratings. The improvement from PI alone (0.647) to full fusion (0.735) demonstrates that behavioral dimensions and ATS features carry complementary signal that the PI type does not capture.

These results must be interpreted with caution. The PI-enhanced models use a small sample (n=229 evaluable), and the confidence intervals for individual PI types are wide (Table 3). However, the AUC=0.735 for full fusion is computed with 5-fold stratified cross-validation on the same sample, providing a reliable point estimate. The finding that PI type is the strongest single predictor—and that it only becomes evaluable through the decision trace connecting PI assessments to HRIS production outcomes—is directionally robust because the effect magnitude is large, though confidence intervals for AUC on this sample would be wide.

Among the six behavioral dimensions, leadership shows a non-linear relationship with production. Production rates by leadership score quartile are: Q1=28.4%, Q2=34.1%, Q3=36.5%, Q4=35.9%—an inverted-U where moderate leadership drive outperforms extreme leadership. For insurance sales specifically, agents with extreme leadership scores may resist the structured coaching and scripted processes that the carrier's training program requires—a pattern consistent with research on curvilinear effects of assertiveness and dominance on team performance (Ames & Flynn, 2007).

## 5.3 Speed-to-Production as a Measurable Economic Constant

Among 679 producing agents in the 2025 deployment cohort, we observe a linear relationship between days-to-SNA and annual premium credit: each day faster to the SNA sales threshold corresponds to $54.35 more in APC (r=−0.258, p<0.000001, 95% bootstrap CI [$39.13, $69.75]). The regression equation is:

$$APC = \$13{,}964 - \$54.35 \times days\_to\_SNA$$

This relationship does not exist in the pre-deployment cohort. Among 1,011 pre-2025 producers with valid APC data, the correlation between speed and production is non-significant (r=−0.032, p=0.31). However, this null finding must be interpreted with caution: the pre-2025 APC data ("year-to-date") is available only for agents still actively producing at data export, meaning 60% of pre-2025 producers have zero APC in the dataset. The null correlation may reflect survivorship-biased data rather than a genuine absence of the speed-production relationship in the historical cohort. The comparison of ramp speed between cohorts is more robust: the median days-to-SNA is 62 days for 2025 producers versus 109 days for pre-2025 producers, a difference of 47 days (Cohen's d=−1.683, KS test p<0.001).

The speed constant survives every robustness check:

| Method | Slope ($/day) | r or p |
|---|---|---|
| Original (n=679) | −$54.35 | r=−0.258 |
| Bootstrap median (10K) | −$54.11 | — |
| Bootstrap 95% CI | [−$69.75, −$39.13] | — |
| Winsorized (1/99%) | −$50.94 | r=−0.259 |

| Method | Slope ($/day) | r or p |
|---|---|---|
| Trimmed (5/95%) | −$31.51 | r=−0.234 |
| Multivariate (source + tenure) | −$34.95 | p<0.0001 |
| Log(APC) transform | slope=−0.0047 | r=−0.276 |
| Quantile P25 | −$12.31 | r=−0.93 |
| Quantile P50 | −$22.27 | r=−0.92 |
| Quantile P75 | −$58.62 | r=−0.97 |

*Table 4. Speed-to-production constant under multiple robustness specifications. Quantile regression r values represent correlation between bucket-level fitted and observed values, not individual-level Pearson r.*

The conservative estimate (multivariate, controlling for source channel and months of active selling) is $34.95/day (p<0.0001, $R^2$=0.180). The relationship strengthens at higher APC percentiles: the P75 quantile slope (−$58.62/day, r=−0.97) is nearly five times the P25 slope (−$12.31/day), indicating that speed acceleration disproportionately benefits higher producers.

The speed constant produces a monotone staircase in tenure-normalized APC across speed buckets:

| Speed Bucket | n | APC/Month (Median) | APC/Month (Mean) | Total APC (Median) |
|---|---|---|---|---|
| 0–30 days | 51 | $1,160 | $1,367 | $9,077 |
| 31–60 days | 239 | $913 | $1,109 | $9,038 |
| 61–90 days | 160 | $856 | $1,005 | $8,710 |
| 91–120 days | 191 | $677 | $744 | $6,995 |
| 121+ days | 38 | $650 | $716 | $6,755 |

*Table 5. Tenure-normalized APC by speed-to-SNA bucket (2025 producers, n=679).*

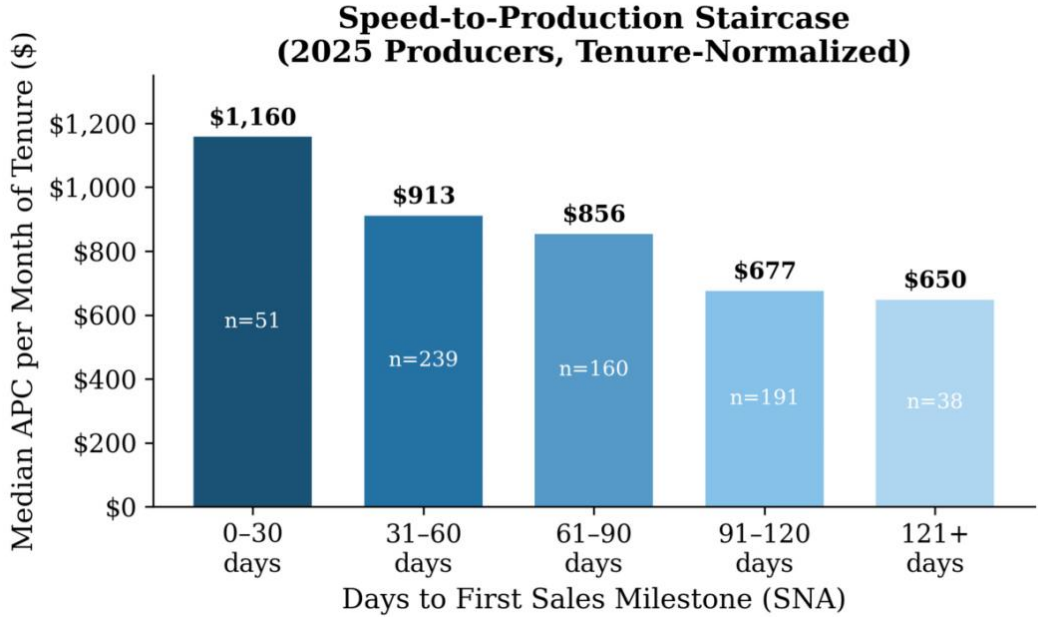

*Figure 2. Speed-to-production staircase showing monotone decrease in tenure-normalized APC by speed bucket.*

The fastest agents (0–30 days) earn $1,160/month—nearly twice the $650/month of the slowest (121+ days). This staircase survives tenure normalization, confirming that the speed-production relationship is not an artifact of differential observation time. Our initial analysis had estimated the speed constant at $56/day; the corrected figure of $54/day after applying proper month-level SNA dates to the full 679-agent population gave us confidence that the underlying relationship was real, not an artifact of the smaller 414-agent subset we had started with.

### 5.4 Behavioral Score as a Moderator of Speed Economics

A critical finding is that the behavioral score does not predict who reaches SNA quickly. Using a median split at score 70: 42.2% of low-scored producers (<70) versus 43.1% of high-scored producers (≥70) reach SNA within 60 days—virtually identical. However, the score does predict who *benefits most* from ramping quickly. To examine the dose-response pattern with finer resolution, we compute the speed constant across three score bands (Table 6; n=668 producers with valid scores, 11 with missing scores excluded):

| Score Band | n | Speed Constant ($/day) | r | p |
|---|---|---|---|---|
| Low (<60) | 109 | −$41.10 | −0.256 | 0.007 |
| Mid (60–80) | 488 | −$46.00 | −0.222 | <0.0001 |
| High (80+) | 71 | −$113.83 | −0.383 | 0.001 |

*Table 6. Speed-to-production constant by behavioral score band (2025 producers, n=668 with valid scores).*

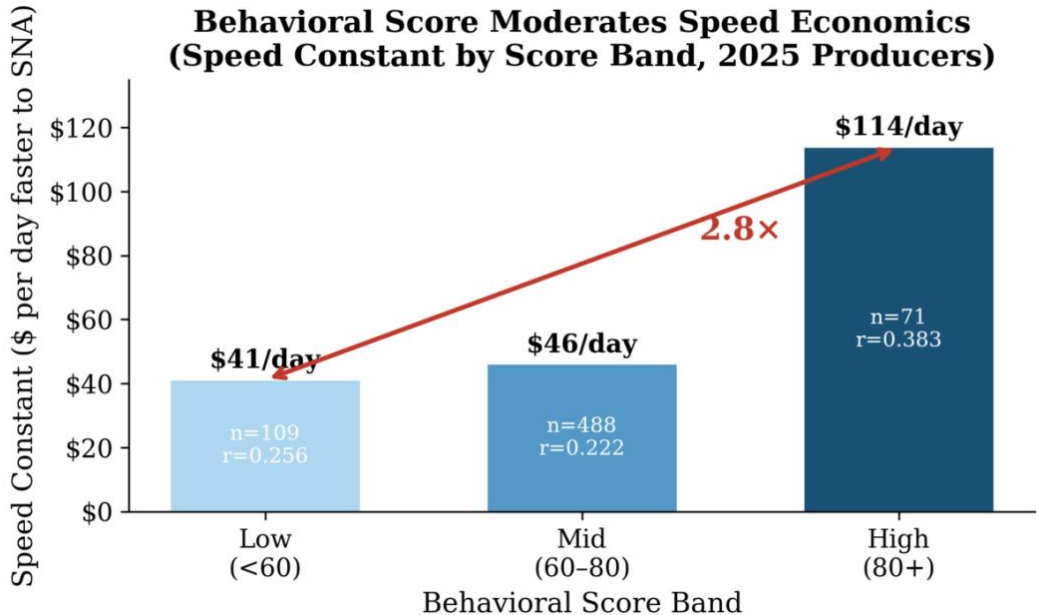

*Figure 3. Behavioral score moderates the speed-production relationship. High-scored agents capture 2.8× the economic value per day of speed acceleration.*

High-scored agents capture $114/day from speed acceleration—nearly three times the $41/day captured by low-scored agents. In practical terms: a high-scored agent who ramps in 30 days versus 120 days gains approximately $10,200 in annual production. A low-scored agent gains approximately $3,700 from the same acceleration. The behavioral score is not a direct predictor of production (AUC=0.57 for deployed initial_fit_score on 714 evaluable 2025 agents)—it is a *moderator* of the speed-production relationship. It identifies agents who convert ramp speed into production most efficiently. The high-score band (80+) has n=71, and the key comparison cells (high-scored fast ≤60d: n=31; high-scored slow >90d: n=15) are small; this dose-response pattern should be considered suggestive pending validation on larger samples.

This moderating effect is also visible in the tenure-normalized data. Among high-scored fast agents (≤60 days), median APC per month is $1,160. Among high-scored slow agents (>90 days), it drops to $589—a 2.0x gap. Among low-scored agents, the corresponding fast/slow gap is only 1.3x ($872 versus $659). The behavioral score amplifies the economic return of speed.

This finding has an important implication for how the scoring system should be evaluated. Traditional classification metrics (AUC for binary production) underestimate its value because the system's primary contribution is not predicting who will produce, but predicting who will benefit most from conditions that accelerate production. A system that identifies candidates whose production scales with speed is strategically different from a system that identifies candidates who will produce at all—even if the AUC for the binary outcome is modest. Nevertheless, the deployed score does exhibit a meaningful calibration curve (Figure 4): production rates nearly double from 16.0% in the lowest score quintile to 31.0% in the third quintile, flattening at higher quintiles. This flattening may reflect restriction-of-range effects (managers preferentially hiring high-scored candidates, compressing the observable score distribution at the top), a genuine ceiling in the score's discriminative power above score ~73, or both.

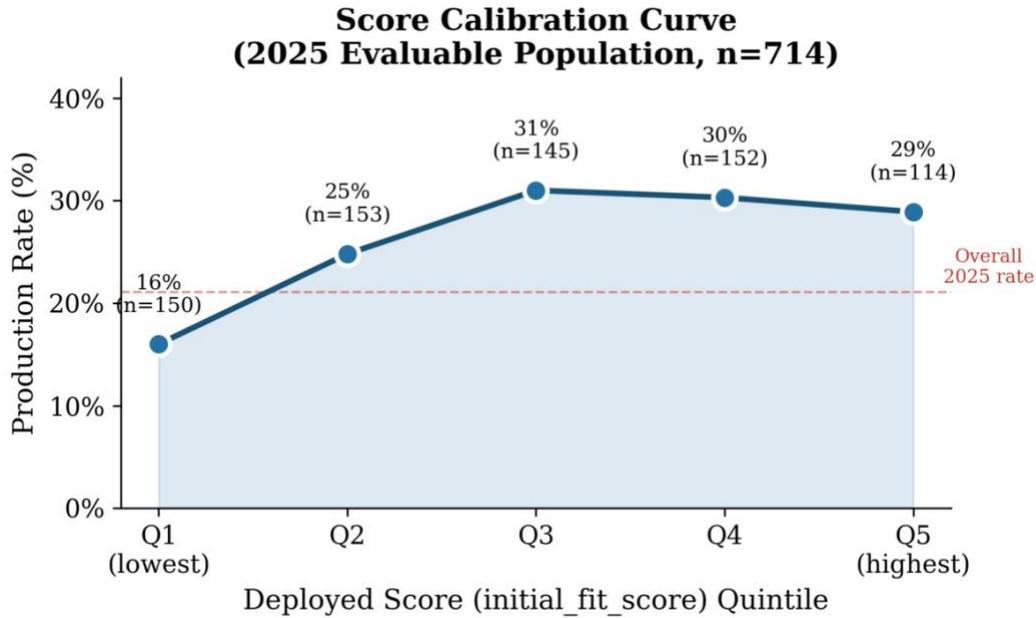

*Figure 4. Score calibration curve for the deployed behavioral score (initial_fit_score) on 2025 evaluable agents (n=714). Dashed line indicates overall 2025 production rate.*

## 5.5 Process-Level Effects: The Phase Transition in Speed-to-SNA

In the pre-deployment cohort (2022–2024), zero producing agents reached SNA within 30 days using day-level HRIS records. In the deployment cohort (2025), 51 agents reached SNA within 30 days and 177 within 60 days using the same HRIS-based measurement. (Month-level SNA dates, which introduce approximately 15 days of noise, show 290 deployment agents within 60 days and 35 pre-deployment agents in the 31–60 day range; we use the more conservative HRIS figures throughout this section.) This is not an incremental improvement—it is a qualitative change in the distribution of ramp times.

This phase transition is not directly attributable to the behavioral scoring system at the individual level. Low-scored and high-scored 2025 agents ramp at virtually identical rates (42.2% versus 43.1% in the 0–60 day bucket). We initially expected the scoring system to explain the speed acceleration—our early analyses attributed it entirely to candidate selection. The within-cohort data forced a different interpretation. The acceleration applies uniformly across score bands, suggesting a *process-level* effect: new hiring patterns, different candidate pools, changed onboarding, or altered lead assignment. The data is consistent with a process-level change in how the organization hires, not just whom it hires.

This distinction is important for understanding what decision trace infrastructure produces. Individual-level prediction (will this candidate succeed?) is the standard framing for hiring AI. But the data suggests the infrastructure's primary effect may operate at the organizational level: by making the relationship between screening inputs and production outcomes visible, the system enables process changes that would not have occurred without the cross-system visibility. The behavioral scoring system may have indirectly enabled this process change by providing confidence to hire non-traditional candidates, but the acceleration itself is an organizational learning effect.

What the scoring system *does* contribute at the individual level, as documented in Section 5.4, is moderating the economic value of this process-driven speed acceleration. The phase transition created a new population of fast-ramping agents. The scoring system identifies which of those agents will convert fast ramp-up into the highest production.

## 5.6 The Speed-Volume Trade-Off: Fewer but Faster Producers

The overall production rate (proportion of hired agents who achieve SNA) declined from 41.5% in 2022 to 21.1% in 2025, after applying proper tenure-based censoring. This decline is concurrent with a more-than-doubling of hiring volume (1,486 agents in 2022 versus 3,623 in 2025) and a significant shift in source channel composition (the dominant job board grew from 44% to 67% of hires). The decline has multiple potential causes that cannot be fully disentangled: the carrier went deeper into the talent pool, market conditions may have shifted, and the scoring system itself may have selected for speed over probability of production.

However, among agents who do produce, tenure-normalized productivity increased substantially. Median APC per month of tenure for 2025 producers is $812, compared to $382–$529 for 2022–2024 producers by year (pooled pre-2025 median: $395). This comparison carries two offsetting caveats. Pre-2025 APC data ("year-to-date APC") is only available for agents still actively producing at data export (survivorship bias inflating the control). Conversely, tenure normalization mechanically favors 2025 agents who have had less time and thus divide a smaller total by a smaller denominator. The direction and magnitude of net bias is unclear; the effect size (Cohen's d=0.443 on these biased samples) should be interpreted as approximate.

A seasonal confound also warrants attention. Agents hired in Q1 produce at 36–38% across all years, compared to 24–29% for Q3 hires. Since the 2025 evaluable population is anchored on agents hired early in 2025 (who have had the most observation time), seasonal effects may contribute to variation in production rates across tenure windows. The speed constant analysis (Section 5.3) is conducted within the 2025 cohort and is not affected by cross-cohort seasonal confounds.

The connected data exposes this trade-off quantitatively. The 2025 process produces fewer producers (21.1% versus 41.5%) but each producer ramps faster (median 62 days versus 109 days) and generates more per month ($812 versus $382–$529). To assess the net economic effect: despite

the lower production rate, the higher hiring volume (3,623 versus 1,486) yielded a comparable number of producers who achieved SNA (680 in 2025, of whom 679 have valid speed and APC data for regression, versus 616 in 2022), though direct APC comparison is limited because only 86 of 616 2022 producers retain measurable year-to-date APC at data export. Using the within-2025 speed constant: each 2025 producer generates approximately \$1,643 more in annual premium than under the prior process (conservative speed constant of \$34.95/day × 47 days of acceleration). Across 680 producers, this represents approximately \$1.12M in additional annual production from speed alone. Without the cross-system connection, the carrier would see only the declining production rate without understanding the compensating productivity gain.

## 6. Discussion

### 6.1 The Decision Trace as a Unit of Organizational Knowledge

The findings in this paper are not predictions from a novel algorithm. They are observations that became possible only when screening inputs, assessment signals, and production outcomes were connected across systems. The anti-predictivity of ATS keywords, the primacy of personality type, and the score-moderated speed constant were all latent in the carrier's data for years.

During the first months of deployment, the most common reaction from hiring managers was disbelief at the keyword findings. Several had personally required insurance experience for over a decade. The production data showed their highest-performing agents were disproportionately those without it.

The structural problem—decision inputs recorded in one system, outcomes in another, reasoning in neither—is not unique to hiring. Clinical trials face a nearly identical challenge: enrollment criteria are recorded in one system, treatment protocols in another, and patient outcomes in a third. When a drug fails a Phase III trial, the question "was the drug ineffective or were the enrollment criteria selecting the wrong patients?" requires the same cross-system trace. The difference is that in hiring, the outcome data already exists in production HRIS systems. The infrastructure cost is not data collection—it is data connection.

### 6.2 Why Keywords Fail

The comprehensive keyword analysis shows that the failure is not limited to a few poorly chosen screening categories—it extends across the entire ATS keyword vocabulary. Of 3,597 testable keywords, the median OR is 0.749 and 70% are directionally anti-predictive. This systemic failure has a plausible mechanism beyond the specific case of insurance experience and license. ATS keyword parsing captures *what candidates have done*, not *how they will perform in a specific role*. The skills, certifications, and domain experience that appear on a resume reflect prior environments, not adaptability to a new one.

For insurance experience and license specifically, agents with industry experience may have developed habits, expectations, and compensation anchors from prior roles that do not transfer to the specific sales environment at this carrier. The personnel selection literature documents that domain experience can be both an asset and a liability depending on the degree of environmental similarity between prior and current roles (Chamorro-Premuzic & Furnham, 2010). Experienced agents may be less coachable, less willing to adopt new processes, and more likely to revert to approaches that worked in different contexts.

The finding that *more keywords correlates with worse production* (r=−0.090, p<0.0001) suggests an additional mechanism: resume padding. Agents with longer keyword lists may be optimizing for ATS filters rather than reflecting genuine capability. Alternatively, agents with diverse but shallow experience across many domains may lack the focus and persistence that insurance sales requires. This connects to the diversity-validity literature: if keyword-based screening produces adverse impact on non-traditional candidate populations while simultaneously failing to predict production, removing it improves both fairness and accuracy (Ployhart & Holtz, 2008).

## 6.3 Scoring as Moderation, Not Classification

The finding that behavioral scoring moderates the speed-production relationship rather than directly predicting production suggests a reframing of how hiring AI should be evaluated. The standard evaluation metric—AUC for binary production—asks whether the system can classify who will succeed. Our data suggests the system's value lies elsewhere: in identifying who will benefit most from favorable conditions. A deployed AUC of 0.57 for binary production (initial_fit_score on 714 evaluable 2025 agents) is modest. But a 2.8x amplification of the speed constant ($114/day versus $41/day) represents substantial economic value that the standard metric does not capture.

This distinction between *prediction* and *moderation* may apply to other AI-assisted decision systems. A loan scoring model might not predict who defaults, but might predict who benefits most from specific loan terms. A clinical decision support system might not predict who recovers, but might predict who responds most to specific treatments. The decision trace primitive enables both types of evaluation.

## 7. Limitations

**Single deployment.** All findings come from one Fortune 500 insurance carrier. Generalization to other carriers, industries, or organizational structures requires additional deployments. The decision trace framework is generalizable; the specific findings (keyword anti-predictivity, PI type rankings, speed constant) may be domain-specific.

**Production rate decline.** The 2025 cohort shows a lower production rate (21.1%) than historical cohorts (34.9–41.5%). Multiple confounding factors—doubled hiring volume, changed source channel mix (dominant job board grew from 44% to 67%), possible market condition shifts, and the scoring system itself selecting for speed over probability—cannot be fully disentangled. The between-cohort comparison is quasi-experimental with a historical control, not a randomized trial.

**Phase transition attribution.** The speed acceleration ($0 \rightarrow 177$ agents in the 0–60 day bucket using HRIS day-level dates) is process-confounded: low-scored and high-scored agents ramp at identical rates. We cannot attribute the speed improvement to behavioral scoring specifically. Onboarding changes, lead assignment changes, market conditions, or other 2025 operational factors may be responsible.

**PI sample size.** The AUC=0.735 for full fusion with PI data is based on n=229 evaluable agents. While the effect size is large and the finding is directionally strong, validation on a larger sample is essential before operational reliance.

**Survivorship bias in APC comparison.** Year-to-date APC for pre-2025 agents is only available for those still actively producing at data export. This biases the control group toward long-tenured survivors and may understate the 2025 APC advantage. Conversely, 2025 agents have a mechanical advantage: they have had less time to accumulate APC, but the year-to-date window captures all of their career production.

**SNA dates are month-level.** Production milestone dates in the HRIS are recorded at month granularity, not day. This introduces approximately 15 days of noise in speed-to-SNA calculations and may inflate the number of agents appearing in faster buckets.

**No office-level clustering.** Standard errors do not account for within-office correlation. Agents in the same office share a manager, market, and training environment. Without office-level random effects or clustered standard errors, confidence intervals may be too narrow.

**Restriction of range in 2025 AUC.** Hiring managers used scores to make decisions, disproportionately not hiring low-scored candidates. The within-2025 AUC of 0.57 (initial_fit_score, n=714) is a lower bound on true predictive power—the system looks less predictive precisely because it already filtered out the candidates it predicted would fail.

**No termination data.** The dataset lacks explicit termination dates. We cannot distinguish agents who "never produced" from agents who "produced and then left." Survival analysis by score band is not possible.

**Retroactive scoring temporal leak.** The model used to retroactively score 2022–2024 agents may have been trained on data that includes some of those agents' outcomes. The retroactive AUC of 0.518 on the pre-2025 cohort may be slightly inflated by information leakage. This does not affect the between-cohort intervention comparison.

**ATS parser fidelity.** The keyword analysis depends on the fidelity of the ATS skill parser; systematic misattribution or omission of skills by the parser could contribute to the observed anti-predictivity pattern.

## 8. Ethics and Data Governance

This study was conducted within the carrier's virtual private cloud (VPC) infrastructure with zero data egress. All data processing, model training, and analysis occurred on the carrier's servers. No individual agent data left the carrier's infrastructure at any point during the study. The carrier reviewed and approved publication of aggregate statistical findings. No individual agents are identifiable in the results presented here: all agent-level examples are anonymized, and no combinations of reported attributes (source channel, score band, speed bucket) identify individuals.

The behavioral scoring model does not ingest demographic data (race, gender, age, national origin, disability status, or geographic location). The input restriction architecture is by design: the system cannot discriminate on attributes it never sees. The dataset itself does not contain demographic fields for any agent. Fairness analysis using source channel as a demographic proxy confirms that all major source channels pass the 4/5ths rule at the median score threshold, and score distributions are similar across channels (mean scores ranging from 66 to 68).

The study did not undergo institutional review board (IRB) review as it was conducted by the technology vendor (AI Synapse Inc.) using de-identified operational data under the carrier's data governance agreement. The analysis constitutes secondary use of existing operational records and does not involve intervention on human subjects beyond the normal scope of the carrier's hiring process. We acknowledge the broader ethical considerations of AI-assisted hiring and note that decision traces, by making screening logic auditable against outcomes, may improve accountability relative to opaque keyword-based screening.

## 9. Future Work

Several directions emerge from this study. First, the PI personality type finding (AUC=0.647, full fusion AUC=0.735) requires validation on a larger sample. The current PI-evaluable subset (n=229) is sufficient to establish direction and approximate magnitude, but operational deployment requires replication with n>1,000. The carrier's ongoing PI assessment program will produce this data over the next 12–18 months.

Second, multi-carrier deployment would test whether the specific findings (keyword anti-predictivity, PI type rankings, speed constant magnitude) are domain-general properties of insurance agent hiring or carrier-specific artifacts. The decision trace framework is designed to be carrier-agnostic; the empirical patterns may not be.

Third, once termination data becomes available, survival analysis by score band would test whether behavioral scoring predicts retention in addition to speed-to-production. If high-scored agents both ramp faster and stay longer, the economic case for the decision trace infrastructure doubles.

Fourth, the moderator finding (Section 5.4) suggests a broader research program: evaluating AI-assisted decision systems not by classification accuracy alone, but by their ability to identify who benefits most from favorable conditions. This reframing—from prediction to moderation—may apply to clinical decision support, credit scoring, educational placement, and other domains where the interaction between individual characteristics and environmental conditions determines outcomes.

Fifth, the qualitative dimension of decision traces—how hiring managers actually interpret and use scores, when they override recommendations and why, and how their mental models of candidate quality compare to what the data shows—remains unexplored. A mixed-methods study combining the quantitative decision traces presented here with structured interviews of hiring managers would substantially strengthen the CSCW contribution by connecting system design to organizational practice.

## 10. Conclusion

This paper presents, to our knowledge, the first empirical study of decision traces deployed at production scale in enterprise hiring. By connecting ATS screening inputs, behavioral assessment signals, and HRIS production outcomes across 10,765 agents at a Fortune 500 insurance carrier, we show that multi-system data fusion surfaces patterns that no single system can detect.

Three findings emerge from the decision trace analysis. First, the signals the industry screens on do not predict production: of 8,181 ATS keywords tested, zero are significantly predictive after multiple comparisons correction, while 30 are significantly anti-predictive. The median keyword is associated with 25% lower odds of production, and keyword-based filtering would reject thousands of productive agents representing millions of dollars in production. Second, personality-based behavioral assessment (Predictive Index) is the strongest single predictor in the dataset, and multi-system fusion reaches AUC=0.735—meaningful predictive power that no individual system approaches. Third, speed-to-production follows a measurable economic constant of $54/day per agent unadjusted, or $35/day controlling for source channel and tenure, moderated by behavioral score such that high-scored agents capture nearly three times the economic value from speed acceleration.

Perhaps most importantly, the behavioral score's value is not in classification but in moderation—identifying agents whose production scales most steeply with ramp speed. This distinction between prediction and moderation has implications beyond hiring: any AI-assisted decision system may benefit from evaluating not just whether it predicts outcomes correctly, but whether it identifies who benefits most from favorable conditions. The decision trace primitive enables both evaluations by connecting inputs to outcomes across systems.

We did not set out to prove that ATS keywords fail. We set out to connect three data systems and see what the connected data showed. What it showed was that the screening signals the industry relies on do not predict the outcomes the industry optimizes for. That finding required no new algorithm—only the infrastructure to ask the question.

## Acknowledgments

The author thanks the carrier's talent acquisition and data teams for their partnership in deploying and evaluating the decision trace infrastructure. The author thanks Hamza Ahmad, Mohsin Ali, and Wassay Ahmed for engineering contributions.

## Conflict of Interest

The author is the founder and CEO of AI Synapse Inc. (NODES), the technology vendor that developed and deployed the behavioral scoring system described in this paper. The carrier independently reviewed and approved publication of aggregate statistical findings. The author has a financial interest in the commercial success of the platform evaluated in this study.

## Data Availability

The data analyzed in this study was provided by the carrier under a confidentiality agreement and is not publicly available. All aggregate statistical results, confidence intervals, and effect sizes are reported in full in this paper. Supplementary materials including detailed robustness checks, fairness analyses, and the full decision log of methodological choices are available upon reasonable request to the corresponding author (saad@nodes.inc).